\documentclass[twoside]{article}

\usepackage{qic}
\usepackage{graphicx} 
\usepackage{subfigure}
\usepackage{bm}
\usepackage{amsmath}
\usepackage{amscd}
\usepackage{afterpage}
\usepackage{float,times}
\usepackage{epsfig}
\usepackage{theorem}
\usepackage{wrapfig}
\usepackage{picinpar}

\input{Qcircuit}

\renewcommand{\thefootnote}{\fnsymbol{footnote}}  

\DeclareSymbolFont{AMSb}{U}{msb}{m}{n}
\DeclareMathSymbol{\N}{\mathbin}{AMSb}{"4E}
\DeclareMathSymbol{\Z}{\mathbin}{AMSb}{"5A}
\DeclareMathSymbol{\R}{\mathbin}{AMSb}{"52}
\DeclareMathSymbol{\Q}{\mathbin}{AMSb}{"51}
\DeclareMathSymbol{\I}{\mathbin}{AMSb}{"49}
\DeclareMathSymbol{\C}{\mathbin}{AMSb}{"43}

\def\R{\mathbb{R}}
\def\N{\mathbb{N}}
\def\C{\mathbb{C}}
\def\Z{\mathbb{Z}}

\newcommand{\D}[2]{\frac{\partial #1}{\partial #2}} 

\begin{document}

\setlength{\textheight}{8.0truein}    

\runninghead{Error Correction Optimisation in the Presence of X/Z Asymmetry} 
            {Z.~W.~E.~Evans, A.~M.~Stephens, J.~H.~Cole and L.~C.~L.~Hollenberg}

\normalsize\textlineskip
\thispagestyle{empty}
\setcounter{page}{1}

\vspace*{0.88truein}

\alphfootnote

\fpage{1}

\centerline{\bf
ERROR CORRECTION OPTIMISATION IN THE PRESENCE OF X/Z ASYMMETRY} 

\vspace*{0.37truein}
\centerline{\footnotesize
Z.~W.~E.~Evans,
A.~M.~Stephens, J.~H.~Cole and L.~C.~L.~Hollenberg}
\vspace*{0.015truein}
\centerline{\footnotesize\it Centre for Quantum Computer Technology,
School of Physics, University of Melbourne}
\baselineskip=10pt
\centerline{\footnotesize\it Melbourne, Victoria 3010,
Australia
}
\vspace*{0.225truein}

\vspace*{0.21truein}

\abstracts{
By taking into account the physical nature of quantum errors it is possible to improve the efficiency of quantum error correction. Here we consider an optimisation to conventional quantum error correction which involves exploiting asymmetries in the rates of $X$ and $Z$ errors by reducing the rate of $X$ correction. As an example, we apply this optimisation to the [[7,1,3]] code and make a comparison with conventional quantum error correction. After two levels of concatenated error correction we demonstrate a circuit depth reduction of at least 43\% and reduction in failure rate of at least 67\%. This improvement requires no additional resources and the required error asymmetry is likely to be present in most physical quantum computer architectures.}{}{}

\vspace*{10pt}

\vspace*{3pt}

\vspace*{1pt}\textlineskip

\setcounter{footnote}{0}
\renewcommand{\thefootnote}{\alph{footnote}}

\section{Introduction}
Quantum computation offers more efficient solutions to some problems which are difficult to solve using any classical computer. These problems include period finding \cite{Shor97}, unstructured searches \cite{Grover96}, and quantum system simulation \cite{Lloyd96, Aspuru-Guzik05}.
To realise these benefits, it is widely accepted that quantum error correction (QEC) will be required to overcome the effects of decoherence and the systematic errors related to the physical implementation of quantum gates. Provided the physical error rate of a quantum computer is below the \emph{fault tolerant threshold}, the fidelity of the computer is improved by QEC. For this reason the fault tolerant threshold is often thought of as the point at which arbitrary accuracy becomes possible through successive levels of error correction. However, with each level of error correction both the physical resources required for the computation and the time spent error correcting increase exponentially. Because of this it is not feasible to use many levels of error correction in a realistic quantum computer. It is desirable to use as few levels as possible.

In order to achieve high fidelity quantum computation with only a few levels of error correction it is important to make optimisations to the error correction process whenever possible. Such optimisations include the design of more efficient error correction codes and circuitry, for example \cite{DiVincenzo07, Bacon06, Ioffe07, Stephens07}, and computer architectures which are specially designed for high performance in the first levels of error correction \cite{Fowler07}.

In this paper we introduce an optimisation to standard QEC which involves exploiting asymmetries in the rates of $X$ and $Z$ errors. In realistic physical systems, the ratio of single qubit $X$ and $Z$ error rates, $\alpha$, is likely to be large. This asymmetry in the error rates can be met with \emph{asymmetric correction of errors} (ACE) to improve efficiency. This technique is of particular interest because it can be applied to most architectures and codes without any extra requirements or alterations.

Section \ref{sec: asymmetric errors} is a brief introduction to QEC, and the origin of the $X$/$Z$ asymmetry. Section \ref{sec: asymmetric correction} describes the workings and limitations of the ACE method. Section \ref{sec: analysis} gives the results of applying ACE to the error correction circuits of the [[7, 1, 3]] code as an example.

\section{Quantum error correction and decoherence asymmetry}
\label{sec: asymmetric errors}
\noindent
In QEC, the information of each logical qubit is represented by a codeword state made up of several physical qubits. Algorithmic gates in the circuit are then replaced with `logical gates' which perform the corresponding operation on the codeword states.
Error correction circuitry is inserted between each logical gate. Each section of circuit which starts and ends on neighboring error correction blocks is known as an `extended rectangle' (an example is shown in Fig.~\ref{fig: extended rectangles}).
The error correction circuits and the logical gate operations are designed according to the rules of fault tolerance \cite{Aliferis06a} so that any single error within each extended rectangle can be tolerated and corrected.
Therefore, a sufficient condition for the success of the logical circuit is that there is at most one error in each of the extended rectangles. Further details of the workings of QEC and proof of the existence of the fault tolerant threshold can be found in \cite{Aliferis06a, Shor96, Knill96, Gottesman98, Preskill98}.

Although decoherence can cause an arbitrary error in the state of a qubit, such an error can be described as a quantum superposition of no error, a bit flip error ($X$), a phase flip error ($Z$), and both a bit and a phase flip error ($Y$). In QEC an arbitrary error on a physical qubit is corrected by performing specific operator measurements to project any deviations from the codeword states into a discrete combination of $X$ and $Z$ errors. The probability of projecting the system into a state affected by an $X$ error or a $Z$ error is related to the amplitudes in the superposition.
For CSS codes, the full error correction circuit can be split into an $X$ correction block and a $Z$ correction block \cite{Steane97}. The $X$ and $Z$ correction circuits are completely independent and so $X$ and $Z$ errors can be treated independently. A distance-3 CSS code can correct up to one $X$ error and one $Z$ error.

\begin{figure}[tb!]
\[
\Qcircuit @C=0.7em @R=1.3em
{
    & \gate{EC} & \gate{H}
    & \gate{EC} & \ctrl{1}
    & \gate{EC} & \qw \\
    & \gate{EC} & \qw
    & \gate{EC} & \targ
    & \gate{EC} & \qw
    \gategroup{1}{2}{1}{4}{0.5em}{.}
    \gategroup{1}{4}{2}{6}{0.9em}{.}
    \gategroup{2}{2}{2}{4}{0.5em}{.}
}
\]
\fcaption{\label{fig: extended rectangles} `Extended rectangles' of QEC are shown by the dotted lines for an example circuit (Bell state preparation). Each line represents a logical qubit, which consists of several physical qubits. The EC boxes represent a complete error correction circuit. The circuit is able to tolerate one single qubit error within each of the extended rectangles.}
\end{figure}

Under the Born and Markovian approximations, decoherence of a single qubit can be modeled by a master equation in Lindblad form \cite{Ike&Mike, Gardiner92, Gardiner04},
\begin{align}
\begin{aligned}
\D{\rho}{t} = \frac{i}{\hbar}[H, \rho] + \frac{1}{2}\sum_{j}(2 L_j \rho L_j^\dagger - L_j^\dagger L_j \rho - \rho L_j^\dagger L_j),
\label{eq: lindblad master}
\end{aligned}
\end{align}
where $\rho$ is the density matrix of the qubit. The primary modes of decoherence are described by $L_1 = \sqrt{\frac{1}{T_1}}\sigma_-$ and $L_2 = \sqrt{\frac{1}{T_2}}\sigma_z$, the Lindblad operators which correspond to population relaxation and dephasing respectively and have characteristic time scales given by $T_1$ and $T_2$. To focus on the effects of decoherence alone we set $[H, \rho] = 0$.

The solution to Eq.~\ref{eq: lindblad master} can be written as an operator sum expansion in terms of the set of Pauli operators and the identity, $E = \{I, X, Y, Z\}$,
\begin{equation}
\rho(t) = \sum_{n,m}{p_{nm}(t)E_n^\dagger \rho_0 E_m}.
\label{eq: lindblad solution}
\end{equation}
The relevant coefficients in this expansion are $p_{xx}$, $p_{yy}$, and $p_{zz}$, which for the decoherence mechanisms described are given by
\begin{equation}
p_{zz}(t) = 1/4(1+e^{-t/T_1}-2e^{-t/(2T_1)-2t/T_2})
\end{equation}
and
\begin{equation}
p_{xx}(t) = p_{yy}(t) = 1/4(1-e^{-t/T_1}).
\end{equation}
These represent the probabilities of projecting the qubit into a state with a $X$, $Y$, or $Z$ error respectively during a cycle of QEC.
Here $t$ is the time of evolution of the system in the presence of decoherence which, for the purpose of QEC, can be thought of as equal to the duration of a physical quantum gate.

For $t\ll T_2 \ll T_1$, the ratio of $Z$ errors to $X$ errors is $\alpha = (p_{zz}+p_{yy})/(p_{xx}+p_{yy}) \approx 2T_1/T_2$.
In Table \ref{tab: T times} we show example values of $T_1$ and $T_2$ for a range of different physical qubit implementations. $T_1$ is typically orders of magnitude longer than $T_2$ and so $\alpha$ is expected to be large.
\begin{table}[tb!]
\begin{center}
\begin{tabular}{c|ccc}
System                  					 & $T_1$     & $T_2$     & $\alpha$ \\
\hline
P:Si \cite{Tyryshkin06}  					 & 1 hour    & 1 ms      & $10^6$            \\ 
GaAs Quantum Dots \cite{Petta05}    & 10 ms     & $>1\mu$s  & $10^4$            \\
Super conducting (flux qubits) \cite{Bertet05}  	 & 4$\mu$s   & 100 ns    & $100$              \\
Trapped ions \cite{Schmidt-Kaler03} & 100ms     & 1ms       & $100$             \\
Solid State NMR \cite{Vandersypen01}& $>1$min   & $>1$s     & $100$              
\end{tabular}
\vspace{10pt}
\tcaption{\label{tab: T times} Approximate $T_1$ and $T_2$ times for a variety of different qubit implementations demonstrating the asymmetry expected between $T_1$ and $T_2$ and hence between $X$ errors and $Z$ errors.}
\end{center}
\end{table}

\section{Asymmetric correction of errors (ACE)}
\label{sec: asymmetric correction}
\noindent
Since $X$ and $Z$ errors are treated independently by QEC, circuit failures can be separated into those caused by excessive $X$ errors and those caused by excessive $Z$ errors. In the presence of an $X/Z$ error asymmetry, $\alpha > 1$, there is a higher risk of failure due to $Z$ errors than failure due to $X$ errors. Therefore, it will be beneficial to focus on the correction of $Z$ errors in order to conserve physical resources and increase overall fidelity.

In conventional QEC, $X$ and $Z$ correction blocks are applied to every logical qubit after every logical gate. 
By removing $X$ correction blocks the circuit is made more vulnerable to $X$ errors, but the effective frequency of $Z$ correction is increased. If circuit failures were previously dominated by $Z$ failures then the overall circuit fidelity will be improved. The removal of $X$ correction blocks will also result in a reduction of the total circuit depth. If multiple levels of error correction are being used this process can be applied to each level of error correction circuitry to further increase circuit fidelity and reduce circuit depth.

\begin{figure}[tb!]
\begin{centering}
\includegraphics[scale=0.35]{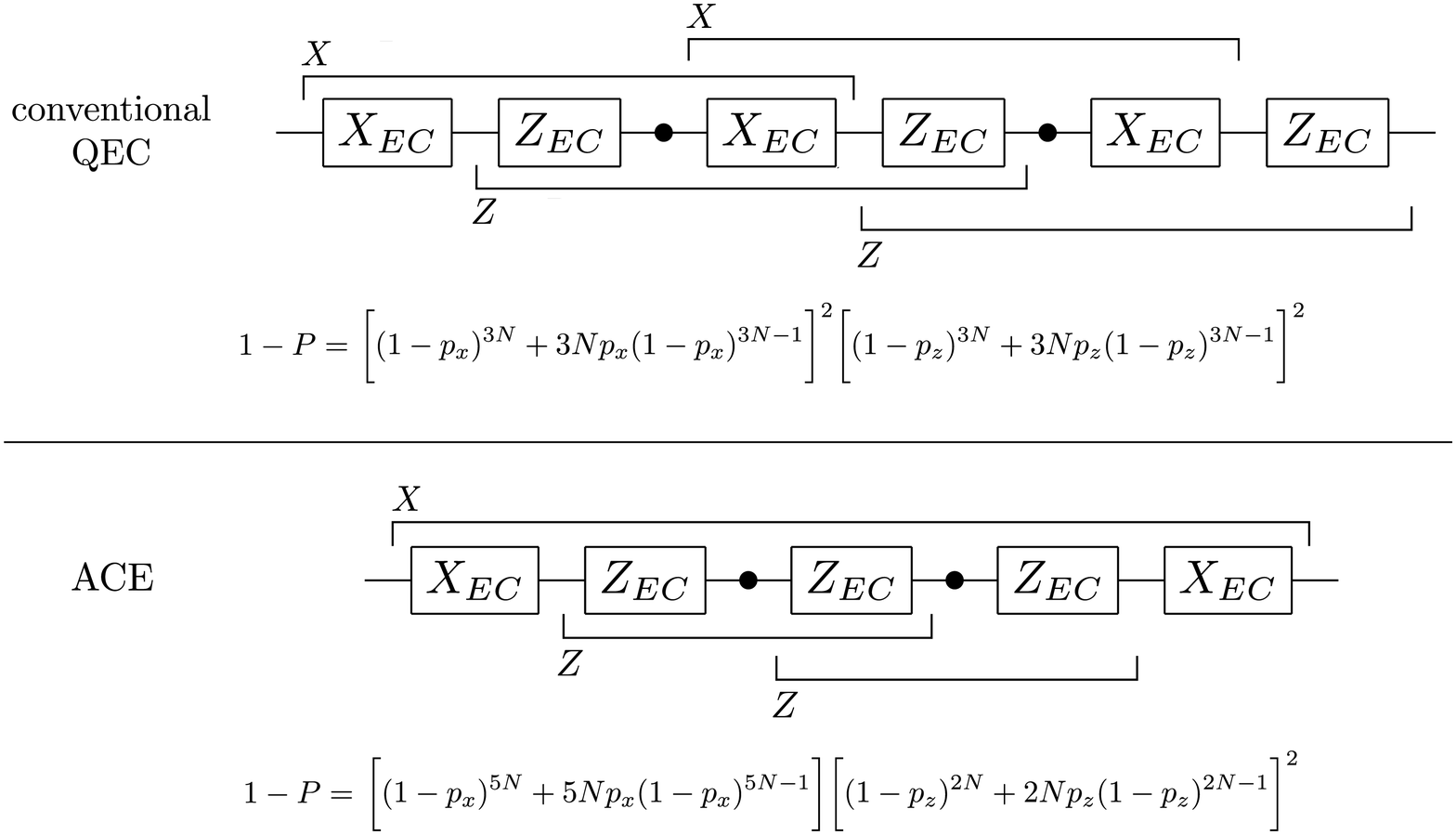}
\vspace{10pt}
\fcaption{\label{fig: ace example1} The upper (lower) braces mark the $X$ ($Z$) type extended rectangles in each of which any single $X$ ($Z$) error will not cause circuit failure. When $X$ error correction is removed the effect is to reduce the circuit depth and raise the $Z$ threshold at the expense of the $X$ threshold. The formulas are approximate expressions for the success rates of the circuits shown. Here $p_x$ is the probability of an $X$ error at each circuit location, $p_z$ is the probability of a $Z$ error, and $P$ is the resulting probability of circuit failure. $N$ is the number of circuit locations in a $X_{EC}$ or $Z_{EC}$ block.}
\end{centering}
\end{figure}

While using this kind of asymmetric correction scheme it is no longer useful to consider the fault tolerant threshold for general errors. In order to analyse the $X$ and $Z$ failure rates separately, the extended rectangles of fault tolerant QEC need to be split into $X$ and $Z$ type extended rectangles (Fig.~\ref{fig: ace example1}). This division of $X$ and $Z$ type extended rectangles must be applied to conventional (non-asymmetric) QEC for comparison with ACE. Using ACE, the $Z$ type extended rectangles will be shorter than those of traditional QEC, and the $X$ type can potentially be much longer. This can be thought of as increasing the $Z$ threshold at the expense of the $X$ threshold.

When implementing ACE, a number of effects must be taken into account. Errors can potentially propagate from one qubit to another through $CNOT$ gates and other two-qubit gates, and so these gates require some special attention. It is still possible to neglect $X$ error correction surrounding $CNOT$ gates, however to retain fault tolerance, we then need to consider super-extended rectangles that include multiple (logical) qubits. Such a rectangle is shown in Fig.~\ref{fig: super extended}. As usual, these rectangles are assumed to fail if there is more than one physical error.
However, any analysis of ACE needs to take into account the possibility that the failure of a super-extended rectangle can result in multiple logical qubit errors.
It is also worth noting that there are usually many possible combinations of two physical errors in a given super-extended rectangle which only result in correctable errors and not circuit failure. Our assumption that any combination of two or more errors will cause a failure assumes the worse case errors and so provides a lower bound for the fidelity.

\begin{figure}
\[
\Qcircuit @C=1em @R=.7em {
    & \gate{X_{EC}} & \gate{Z_{EC}} & \qw
    & \gate{Z_{EC}} & \ctrl{1}
    & \gate{Z_{EC}} & \gate{X_{EC}} & \qw \\
    & \gate{X_{EC}} & \gate{Z_{EC}} & \targ
    & \gate{Z_{EC}} & \targ
    & \gate{Z_{EC}} & \gate{X_{EC}} & \qw \\
    & \gate{X_{EC}} & \gate{Z_{EC}} & \ctrl{-1}
    & \gate{Z_{EC}} & \qw
    & \gate{Z_{EC}} & \gate{X_{EC}} & \qw
    \gategroup{1}{2}{3}{8}{0.9em}{--}
    \gategroup{1}{3}{1}{5}{0.6em}{.}
}
\]
\fcaption{\label{fig: super extended} In circuit regions which allow errors to spread across multiple qubits; extended rectangles must be stretched further to ensure fault tolerance. The dotted box shows a $Z$-type extended rectangle. The dashed box shows an $X$-type super-extended rectangle which can be thought of as $X$ correction surrounding a 3-qubit gate. As usual, these rectangles are assumed to fail if they experience more than one error.}
\end{figure}

Not all quantum gates commute with the quantised $X$ and $Z$ errors. For example the Hadamard gate, $H$, does not commute, $HZ = XH$. Therefore any uncorrected $Z$ errors prior to a Hadamard gate will effectively become $X$ errors and vice-versa. To simplify the error analysis, we always include both $X$ and $Z$ correction on either side of Hadamard gates and other gates with error mixing properties, such as $T$ and $S$ \cite{Ike&Mike}. This limits the number of $X$ correction blocks that can be removed, and so the asymmetry in the errors may not be fully exploited by a single level of error correction. That is, after a single level of asymmetric error correction, the failure rate due to $Z$ errors may still be higher than the failure rate due to $X$ errors.

Aside from positioning $X$ correction blocks surrounding Hadamard gates and other problematic gates, the optimal number of $X$ correction blocks to remove will depend on the asymmetry in the error rates. To gain any benefit from using ACE, the largest of the $X$ extended rectangles should be kept small enough so that the $X$ failure rate does not overtake the $Z$ failure rate. Specific circuitry and error rate information needs to be taken into account in order to find the optimum error correction scheduling. Ultimately, classical software could be used to control this scheduling. The software could potentially analyse error rates in real time to adjust and improve the error correction process via ACE while the quantum computer is running.

\section{Speed and fidelity gains}
\label{sec: analysis}
\noindent
In general, the application of ACE will depend on the specifics of the circuit; including the positions of Hadamards and two-qubit gates. However, for two or more levels of QEC, ACE can be applied to the error correction circuitry itself to improve the speed and fidelity of every logical gate.

\begin{figure}[tb!]
\begin{centering}
\mbox
{
    \subfigure{\includegraphics[scale=0.35]{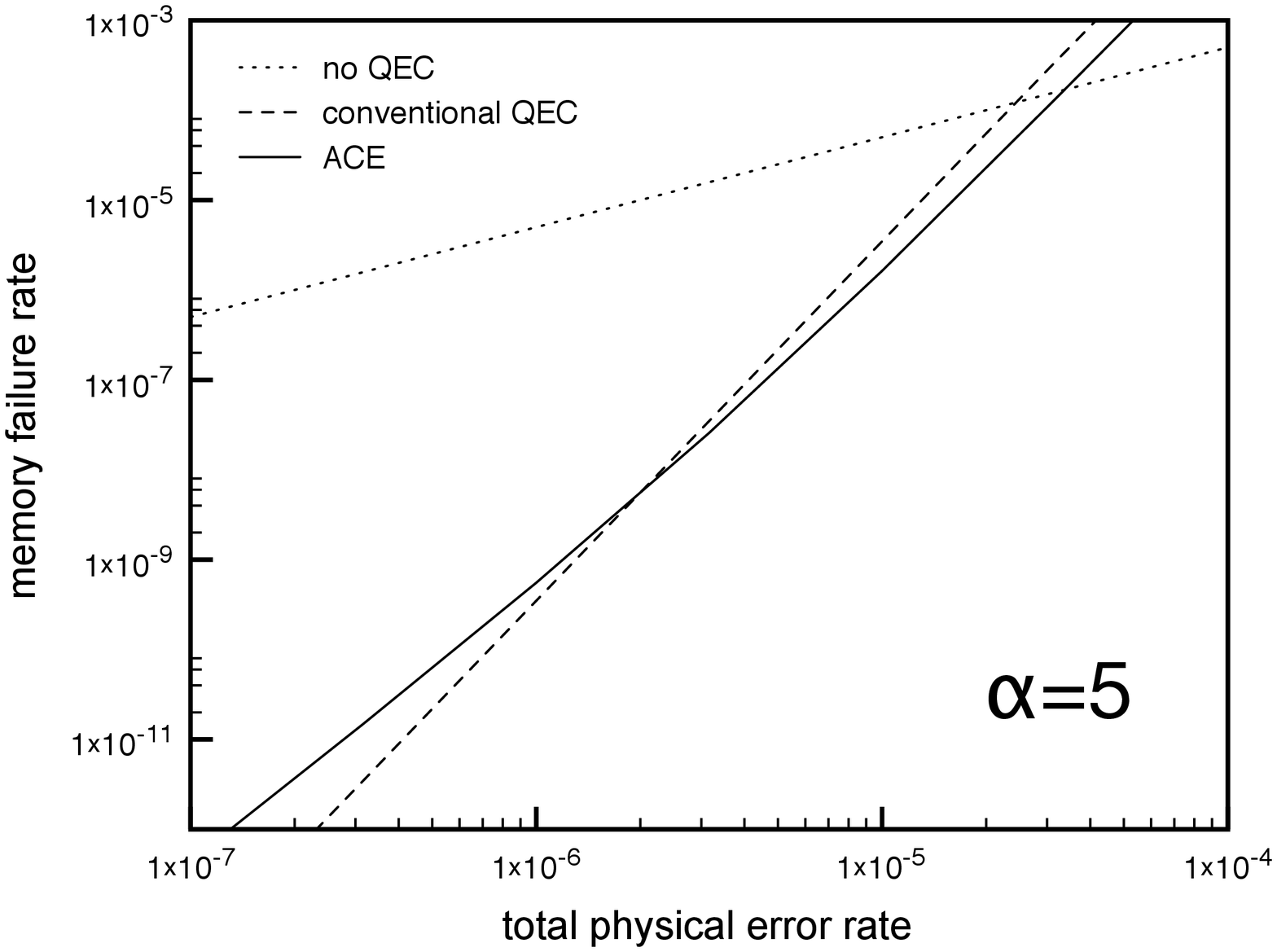}}
    \subfigure{\includegraphics[scale=0.35]{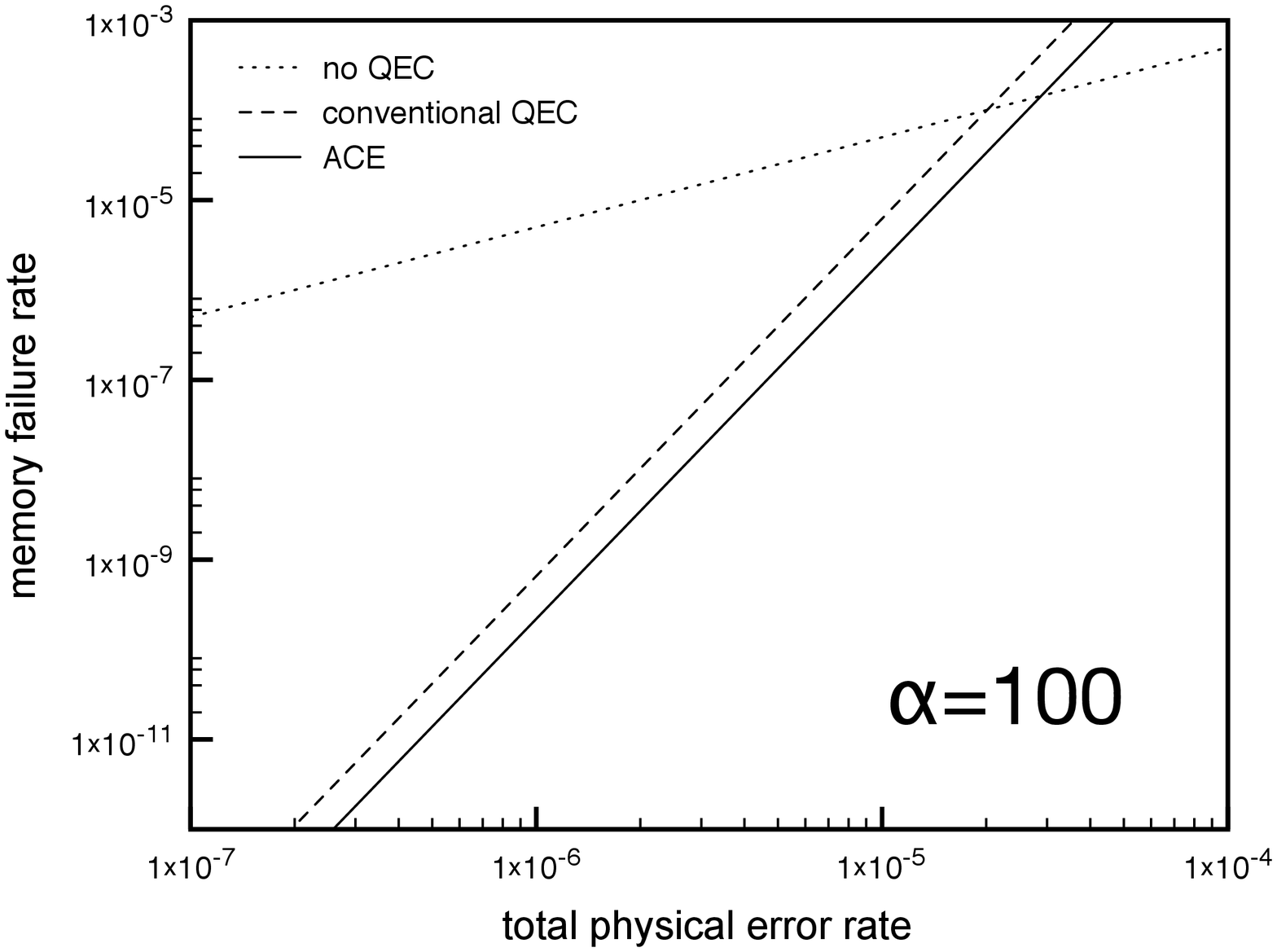}}
}
\vspace{5pt}
\fcaption{\label{fig: results plot} Circuit failure rates at two levels of error correction for a string of five memory locations. Larger asymmetries are required for ACE to gain a fidelity advantage over conventional QEC at lower physical error rates.}
\end{centering}
\end{figure}

Taking the [[7,1,3]] code \cite{Steane96} as an example, and using the ancilla decoding method proposed in \cite{DiVincenzo07}, we have analysed the error correction circuit using a large degree correction asymmetry. For simplicity we assume that non-local gates are available and that all physical circuit locations have equal probability of error. Any $X$ correction blocks not surrounding Hadamard gates were removed or, in cases for which this would disrupt the synchronisation of the circuit, replaced with $Z$ correction. This results in a reduction in the depth of the circuit by around 25\%. Calculations of a lower bound for the fidelity in the modified circuit versus the unmodified circuit show that an asymmetry factor of at least $5$ is required for ACE to gain any benefit in logical fidelity over conventional QEC. For $\alpha \geq 15$ the reduction in the failure rate flattens out to a factor of $2$. The condition that the asymmetry factor be greater than $5$ is dependent on the error correction circuit used, and on the total error rate. These estimates were made using a total error rate of $10^{-5}$. Lowering the total error rate increases the asymmetry at which this ACE implementation becomes preferable to conventional QEC. While these gains are relatively small, they are effectively free. The reduction in circuit depth and increase in fidelity arise only from changes in the order and frequency of the QEC procedure which target the inherent decoherence asymmetry.

It is important to note that using this particular error correction circuit for the [[7,1,3]] code, physical $Z$ errors during the error correction process will never manifest as $X$ errors on the data qubits. This is a vital property of the error correction process to ensure that $X$ errors will not occur with the probability of the $Z$ errors when ACE is applied to the error correction blocks.

If, after one level of asymmetric error correction, the failure rate due to $Z$ errors is still higher than the failure rate due to $X$ errors, then the asymmetry has not yet been fully exploited. However, for the application of higher levels asymmetric correction to be beneficial, there needs to be an asymmetry in the errors in \emph{logical qubits}. The [[7,1,3]] code ensures that physical $Z$ failures cannot produce logical $X$ errors and vice-versa. So for this code, the nature of the error asymmetry is preserved at each level of error correction. This is not true for all codes. Some codes, such as the Bacon-Shor code \cite{Bacon06, Aliferis06b}, reverse the $X/Z$ asymmetry at each level of error correction, and so the asymmetric correction should also be reversed. For other codes the asymmetry may be diminished, or completely removed, by the process of error correction.

Assuming that error correction is representative of a typical quantum circuit, it is reasonable to expect that higher level $X$ correction blocks can be removed to reduce the depth by a further 25\% for each level of error correction. The total reduction in the depth of the circuit will then be 43\% for two levels of error correction and potentially more if more levels of error correction are used. For two levels of error correction we compare conventional error corrected memory with 12 error correction blocks (6 $X$ and 6 $Z$) and 5 waits to a fully ACE modified memory with 9 error correction blocks (2 $X$ and 7 $Z$) and 5 waits, similar to the arrangement in Fig.~\ref{fig: ace example1}. The resulting fidelities are shown in Fig.~\ref{fig: results plot}. For sufficiently large error asymmetry, the ACE method has a $3$ times lower failure rate. For a total error rate of $10^{-5}$, the maximum benefit is reached at an asymmetry factor of around $\alpha = 10$.

A more detailed analysis of ACE may involve taking into account expected types of systematic errors from poorly characterised gates. Also, by considering `hybrid' extended rectangles, which contain a Hadamard gate, it may be possible to draw a greater advantage from ACE by removing additional $X$ correction boxes.

In our analysis of the [[7,1,3]] code we removed around half of $X$ correction, and replaced some of the remainder with $Z$ correction. We have not attempted to find the optimal implementation. For an approximate bound on the theoretical maximum benefit that can be gained from ACE with two levels of error correction, consider a quantum computer which uses no $X$ correction at all. In this case there would be a reduction in circuit depth of around $75\%$ and a reduction in failure rate of around 20 times. Not only would achieving this limit require a large error asymmetry, but we are still limited by the presence of error mixing gates ($T$, $S$, and $H$).

\section{Conclusions}
\label{sec: conclusions}
\noindent
With a finite number of physical qubits, the fault tolerant threshold should not be thought of as the critical point for achieving arbitrary accuracy in quantum computing. With only a few levels of error correction available, all possible improvements in the efficiency of error correction must be considered to reach the high fidelity required for large scale quantum calculations. Asymmetric correction of errors is a simple example of how a greater understanding of quantum errors can lead to such an improvement. Other similar examples may include correcting for errors during memory less frequently than during computation if errors due to decoherence are less likely than systematic errors, or targeting error correction at the errors that arise from the implementation of the most common physical operations. This may be aided by the selective use of characterisation or composite pulses \cite{Testolin07}.

In conclusion, we have shown that asymmetries in $X$ and $Z$ error rates can be exploited to increase the speed and fidelity of any error corrected quantum circuit by neglecting the correction of the less common errors. Doing so requires no additional resources and does not prohibit universal quantum computation, and will be beneficial in the many QC architectures that are expected to exhibit the required error asymmetry.

\nonumsection{Acknowledgments}
\noindent
We would like to thank S. Devitt, A. Fowler and A. Greentree for helpful discussions and suggestions.
This work was supported in part by the Australian Research Council, the 
Australian Government and by the US National Security Agency, Advanced 
Research and Development Activity and the US Army Research Office under 
contract number W911NF-04-1-0290.

\vspace{50pt}

\bibliographystyle{unsrt}
\bibliography{paper}

\end{document}